\documentclass[
 aip,
 cha,
amsmath,
amssymb,
reprint
]{revtex4-1}
\usepackage{graphicx}
\usepackage{dcolumn}
\usepackage{bm}

\usepackage[utf8]{inputenc}
\usepackage[T1]{fontenc}
\usepackage{mathptmx}

\begin{document}


\title{Weak Electric-Field Detection with Sub-1 Hz Resolution at Radio Frequencies Using A Rydberg Atom-Based Mixer}

\author{Joshua A. Gordon}
\affiliation{National Institute of Standards and Technology, Boulder, CO, USA
}%

\email{josh.gordon@nist.gov}

\author{Matthew T. Simons}
\author{Abdulaziz H. Haddab}%
\affiliation{University of Colorado, Boulder, CO, USA
}%

\author{Christopher L. Holloway}
\affiliation{National Institute of Standards and Technology, Boulder, CO, USA
}%

\date{\today}

\begin{abstract}
Rydberg atoms have been used for measuring radio-frequency (RF) electric (E)-fields due to their strong dipole moments over the frequency range of 500 MHz-1 THz. For this, electromagnetically induced transparency (EIT) within the Autler-Townes (AT) regime is used such that the detected E-field is proportional to AT splitting. However, for weak E-fields AT peak separation becomes unresolvable thus limiting the minimum detectable E-field. Here, we demonstrate using the Rydberg atoms as an RF mixer for weak E-field detection well below the AT regime with frequency discrimination better than 1~Hz resolution. Two E-fields incident on a vapor cell filled with cesium atoms are used. One E-field at 19.626000~GHz drives the  $34D_{5/2}\rightarrow 35P_{3/2}$ Rydberg transition and acts as a local oscillator (LO) and a second signal E-field (Sig) of interest is at 19.626090~GHz. In the presence of the LO, the Rydberg atoms naturally down convert the Sig field to a 90~kHz intermediate frequency (IF) signal. This IF signal manifests as an oscillation in the probe laser intensity through the Rydberg vapor and is easily detected with a photodiode and lock-in amplifier. In the configuration used here, E-field strength down to $\approx$~46~$\mu$V/m~$\pm$~2~$\mu$V/m were detected. Furthermore, neighboring fields 0.1~Hz away and equal in strength to Sig could be discriminated without any leakage into the lock-in signal. For signals 1~Hz away and as high as +60~dB above Sig, leakage into the lock-in signal could be kept below -3~dB.
\end{abstract}

\maketitle

Rydberg atoms \cite{gallagher_1994} have been demonstrated as quantum sensors for electric(E)-field metrology over the radio frequency (RF) range of approximately 500~MHz-1~THz, and have properties not found in classical E-field sensors, such as sub  RF-wavelength size \cite{HollowaySubwave, FanSublabmda, fiber_coupled_simons_18,cox_quantumatomreceiver2018}, self calibration \cite{HollowayBroadbandRydberg,subnear_2017}, and system international (SI) traceability to Plank's constant \cite{NIST_SI2018}. Electromagneticlly induced transparency \cite{EITMarangos, adams_microwavedress2011, Mohaptra2007CoherentOptDet}(EIT), and Autler-Townes (AT) splitting \cite{AutlerTownesPR100} used to realize the Rydberg atom E-field sensor, reduce an RF E-field measurement to an optical frequency measurement. Progress has been made using Rydberg atoms to characterize classical properties of RF E-fields including magnitude \cite{HollowayBroadbandRydberg,subnear_2017,Gordon2014MillimeterWaveDet,Sedlacek2012Nature}, polarization \cite{Sedlacek2013AtomBasedVector}, phase \cite{SimonsAtomMixer2019} and, power \cite{HollowayPower2018APL}. More recently the concept of the Rydberg E-field sensor has been expanded in the form of  the “Rydberg Atom Receiver" and "Rydberg Atom Radio" \cite{cox_quantumatomreceiver2018,meyer_digitalcomm2018, song_credibility2018, AtomRadioanderson2018, MIT_atomradio2018, Multibandradio_holloway2018} which have been used to detect time varying fields of common modulation schemes such as QPSK, AM, and FM. \\ 
\indent The detection of weak RF fields (i.e. below 1~mV/m) is important for practical applications if the Rydberg atom RF field sensor is to compete with traditional circuit based sensors. Techniques using optical cavities\cite{CavitySens} to narrow the EIT line width and improve AT splitting resolution, as well as homodyne detection\cite{Kumar_RFsensitivity_SR2017} with a Mach Zehnder interferometer in order to reduce signal to noise levels have been proposed for weak RF field measurements. Some of the weakest RF fields as low as 800~$\mu$V/m  have been detected thus far by fitting models to EIT probe laser specrtra in search of small perturbations\cite{Sedlacek2012Nature}.  Previously we reported on the Rydberg atom mixer \cite{SimonsAtomMixer2019} for determining the phase of an RF field. Here, we show how this mixer effect can be applied for the detection of weak RF fields that are well below AT splitting with the added benefit of isolation of signals at adjacent frequencies, and frequency selectivity of $\sim 10^8$ better than that provided by the Rydberg transition alone. Using the Rydberg atom mixer we demonstrate a lowest detectable field of 46~$\mu$V/m without the need for cavities or inteferometers with better than $\sim 1$~Hz resolution.

\begin{figure}[htbp]
\includegraphics[width = 1\linewidth]{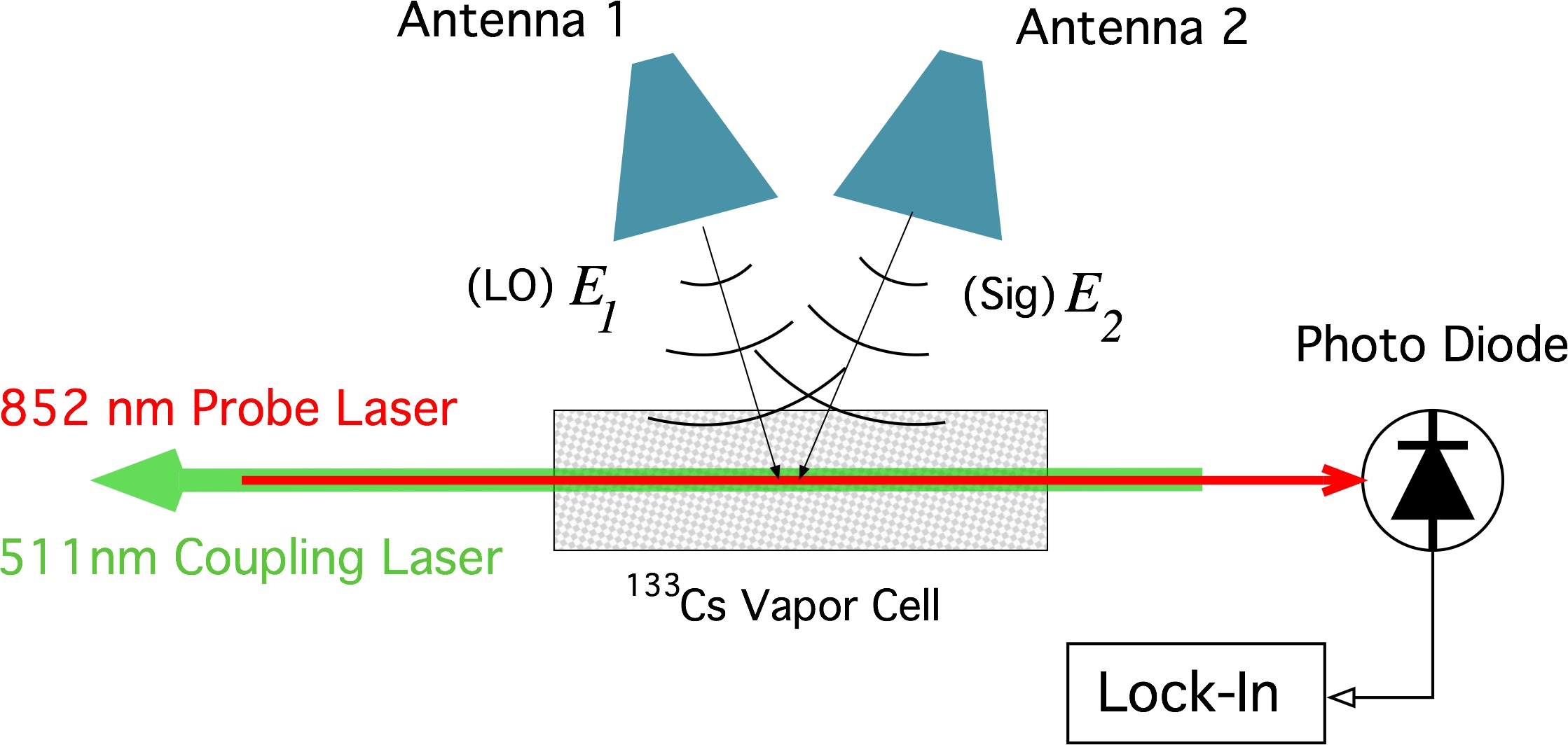}
\caption{\label{fig:setup}Diagram of experimental setup.  E-field  $E_{1}$ acting as a local oscillator (LO) is produced by Antenna 1 while Signal (Sig) E-field $E_{2}$ is produced by Antenna 2. Both fields are superposed along with the probe and coupling lasers at the $^{133}$Cs vapor cell. Probe laser is incident on the photodiode with output passed to a lock-in amplifier. }
\end{figure}	
	
\begin{figure}[htbp]
\includegraphics[width = 0.8\linewidth]{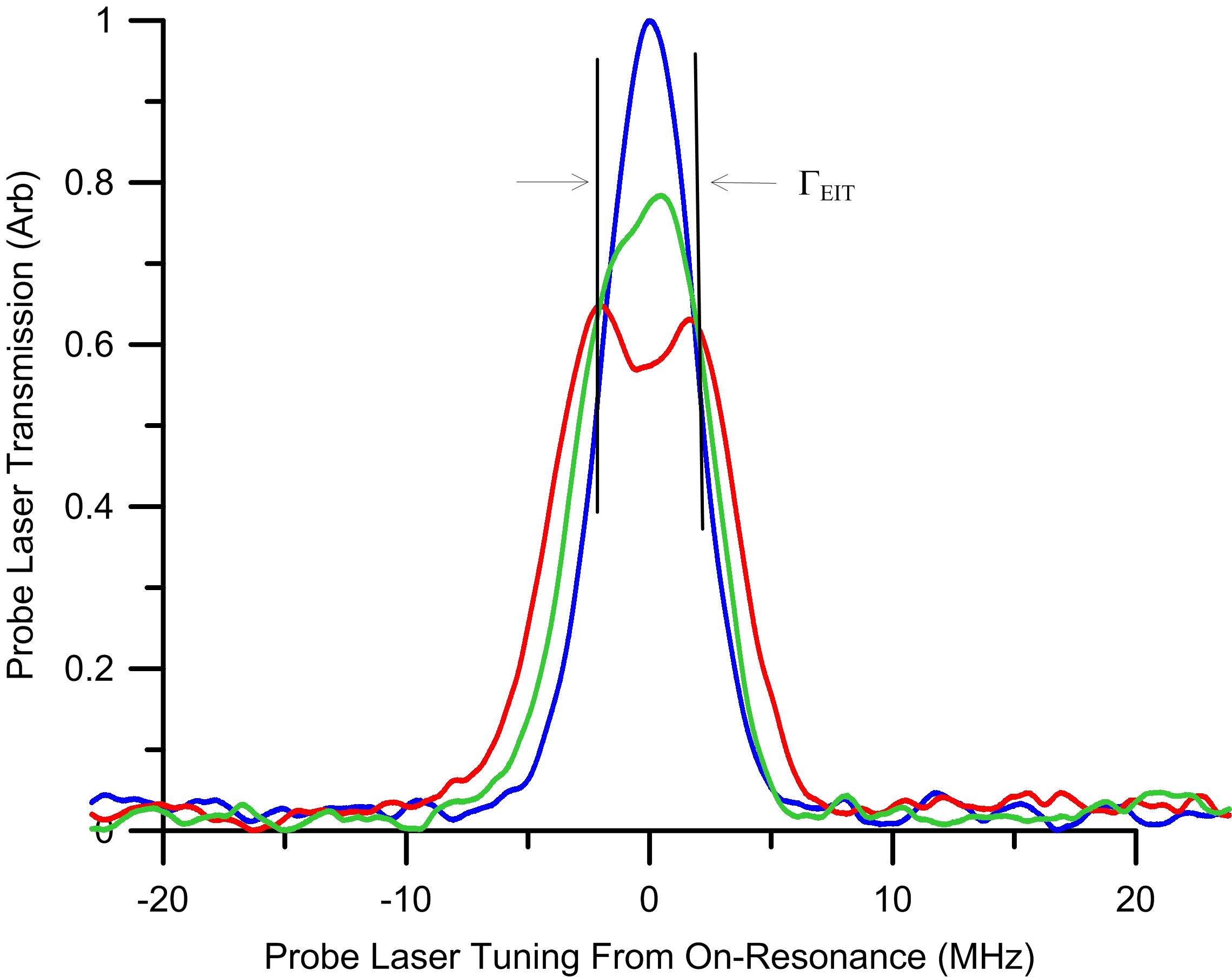}
\caption{\label{fig:EITGamma} Probe laser spectrum plots showing the transition into the AT regimes. (Blue) no RF field where E=0~V/m, (Green) $E<~E_{AT}$, (Red) $E=~E_{AT}$ the EIT peak just begins to split into two resolvable peaks separated by $\Gamma_{EIT}$. }
\end{figure}

The setup for this work is shown in Fig.\ref{fig:setup}. Rydberg atoms are produced using a 75~mm$\times$25~mm (Length$\times$Diameter) cylindrical glass atomic vapor cell filled with cesium ($^{133}$Cs) atoms. A probe laser tuned to the the $D2$ transition wavelength of $\lambda_{p}$=852~nm excites the $^{133}$Cs from the ground state to the first excited state ($6S_{1/2}\rightarrow 6P_{3/2}$). A counter propagating coupling laser is tuned to $\lambda_{c}$=511.148~nm, and further excites the $^{133}$Cs atoms to the Rydberg state $34D_{5/2}$.  The coupling laser also acts to produce the EIT in the probe laser. The probe laser beam has a full-width half-maximum (FWHM) of 425~$\mu$m and a power of 49~$\mu$W, the coupling laser has a FWHM of 620~$\mu$m and a power of 60.6~mW. Under these conditions an incident RF field operating near the frequency of 19.626 GHz drives the $34D_{5/2}\rightarrow 35P_{3/2}$  transition. With the probe laser frequency fixed on resonance with the D2 transition, the transmission through the vapor cell is in general reduced when in the presence of the applied RF field. For appreciable field strengths the atoms are driven to the Autler-Towns regime \cite{AutlerTownesPR100} which splits the observed EIT peak in the probe laser transmission spectrum. The frequency separation $\Delta f_{AT}$ of the two AT peaks is given \cite{HollowayBroadbandRydberg,Sedlacek2012Nature} by,
\begin{equation}
\Delta f_{AT}=\frac{\lambda_{c}}{\lambda_{p}}\frac{E_{RF}\wp_{RF}}{2\pi\hslash}
\label{eq:deltaf}
\end{equation}
Where $\wp_{RF}$ is the dipole matrix element of the RF Rydberg transition and $\hslash$ is Plank's constant. The dipole moment for the resonant RF transition is $\wp=723.3739 e a_0$ (which includes a radial part of $1476.6048 e a_0$ and an angular part of $0.48989$, which correspond to co-linear polarized optical and RF fields, where $e$ is the elementary charge; $a_0=0.529177\times 10^{-10}$~m and is the Bohr radius). AT splitting as a method for E-field sensing becomes less effective for E-fields too weak to cause resolvable AT peak separation. The work described below overcomes this weak E-field limitation through the Rydberg atom mixer effect with the added benefit of narrow band frequency selection and tuning. Here, we define the minimum detectable RF field capable of being detected with AT splitting as that which causes an AT peak separation equivalent to the EIT line width $\Gamma_{EIT}$. From (\ref{eq:deltaf}) this is, 
\begin{equation}
E_{AT}=\frac{\lambda_{p}}{\lambda_{c}}\frac{2\pi\hslash~\Gamma_{EIT}}{\wp_{RF}}.
\label{eq:emin}
\end{equation}
As determined from the EIT spectrum shown in Fig. \ref{fig:EITGamma}, $\Gamma_{EIT}\approx$~4~MHz and $E_{AT}$=0.72~V/m for the above mentioned Rydberg states. 

\begin{figure}[htbp]
\includegraphics[width = 0.8\linewidth]{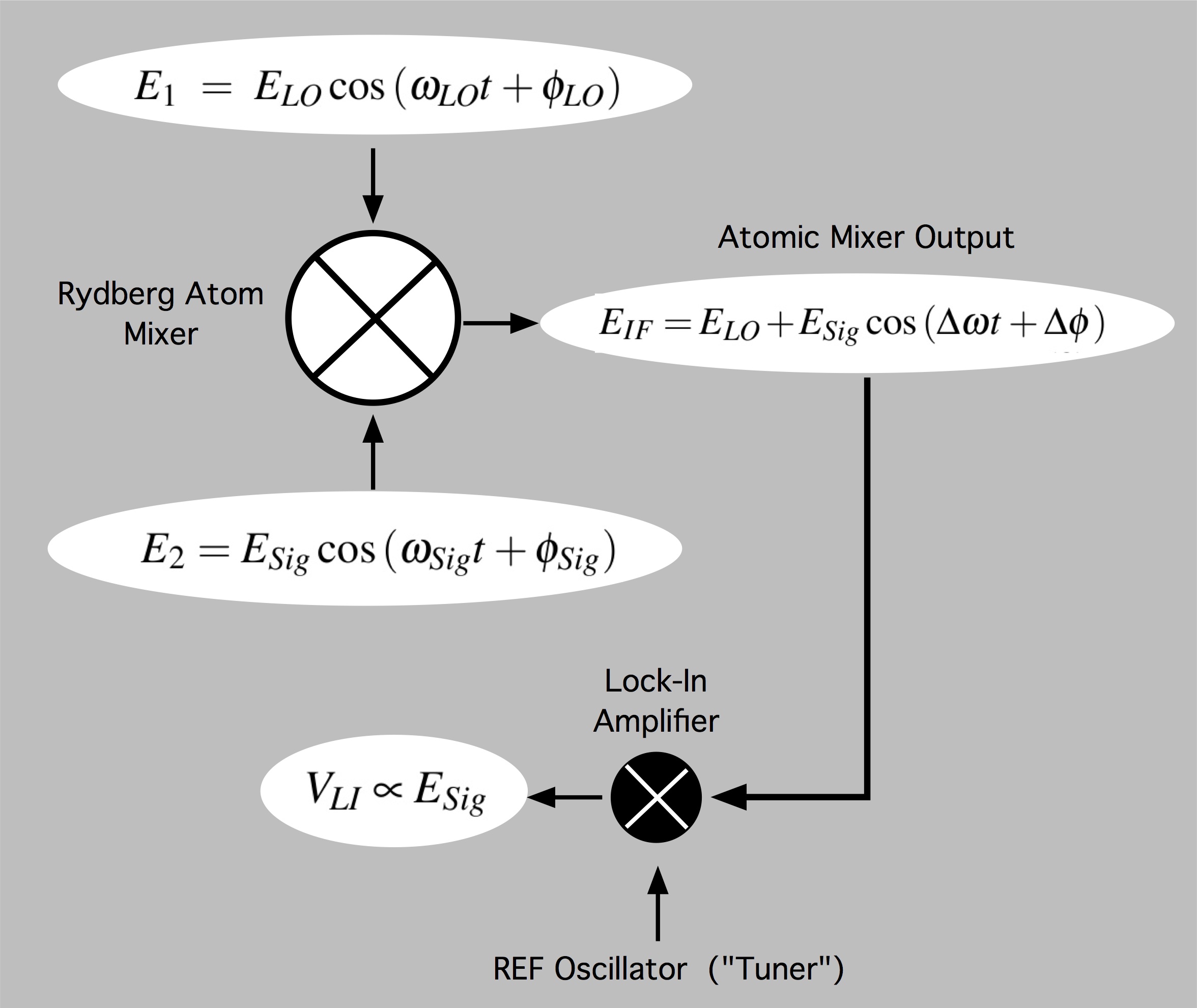}
\caption{\label{fig:flow}Flow chart showing the application of the Rydberg atom mixer to weak field detection. Inputs Local Oscillator (LO) field $E_{1}$ and Signal (Sig) field $E_{2}$, produce an IF field $E_{IF}$ output  which is detected by a lock-in amplifier producing a voltage $V_{LI}$ proportional to $E_{Sig}$. The reference oscillator (REF Oscillator) for the lock-in is set to $f_{REF}$=$f_{IF}$ and allows narrow band $\lesssim$ 1~HZ selective tuning to isolate Sig fields having a range of differing frequencies relative to the LO frequency.}     
\end{figure}

\begin{figure}[htbp]
\includegraphics[width = 0.9\linewidth]{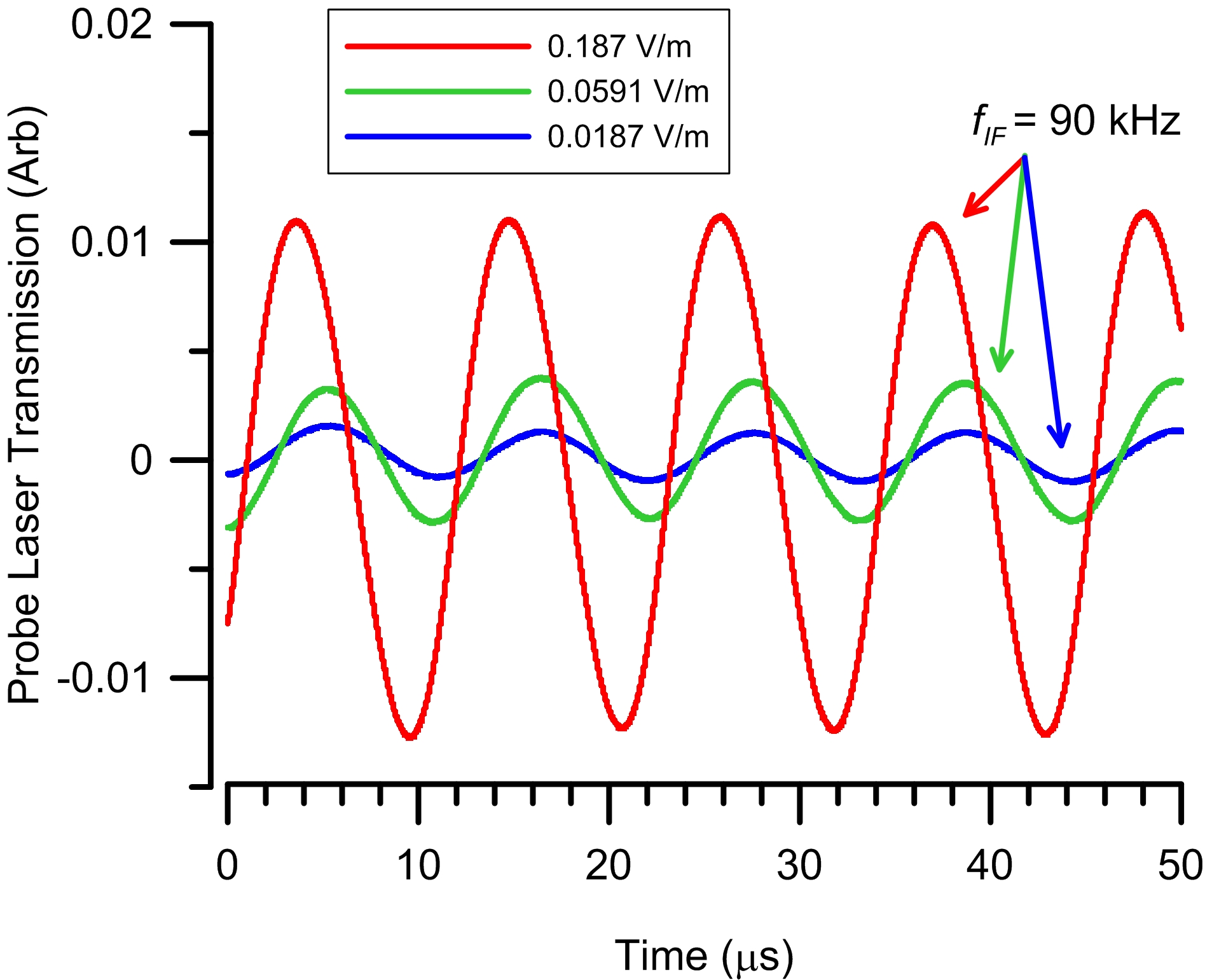}
\caption{\label{fig:IFsig}Time domain plots of the IF signal from the photodiode for $E_{Sig}$=0.187~V/m, 0.0591~V/m, 0.0187~V/m}
\end{figure}

A schematic of the Rydberg atom mixer\cite{SimonsAtomMixer2019} is shown in Fig.~\ref{fig:flow}. Two different RF fields are incident on the vapor cell, $E_{1}=E_{LO}\cos\left(\omega_{LO}t+\phi_{LO}\right)$, and $E_{2}=E_{Sig}\cos\left(\omega_{Sig}t+\phi_{Sig}\right)$. One is tuned to $f_{LO}$=$\omega_{LO}/{2\pi}$=19.626000 GHz such that it is on resonance with $34D_{5/2} \rightarrow 35P_{3/2}$  Rydberg transition. This field acts as a local oscillator (LO). The second field $E_{2}$ is the signal field (Sig) that is to be sensed and is tuned to $f_{Sig}$=$\omega_{Sig}/{2\pi}$=19.626090~GHz such that it is detuned by +90~kHz from the LO field. Here, we explore the case when both $E_{1}$ and $E_{2}$ are co-polarized and considered weak where $E_{1}\approx E_{AT}$ and $E_{2}\leq E_{AT}$.\\
\indent The interference occurring from the superposition of these fields results in a high frequency component $E_{res}$ and low frequency component $E_{mod}$. With $\bar{\omega}=(\omega_{LO}+\omega_{Sig})/2$, $\Delta\omega=\omega_{LO}-\omega_{Sig}$, and $\Delta\phi=\phi_{LO}-\phi_{Sig}$, for small relative detuning where $\Delta\omega / \bar{\omega}\ll1$ the total field at the atoms $E_{atoms}$ can be shown to be, 

\begin{align}
&E_{atoms}= E_{1}+E_{2}\label{etot}\\
&=\cos\left(\omega_{LO} t + \phi_{LO}\right)\sqrt{E^2_{LO}+E^2_{Sig}+2{E_{LO}}{E_{Sig}}\cos\left(\Delta\omega t+\Delta\phi \right)} \label{eq:eatoms} \\ 
&  = E_{\textrm{res}} \times E_{\textrm{mod}}.\label{eq:resmod} 
\end{align}
Where $E_{res}$ oscillates at $\omega_{LO}$ and $E_{mod}$ oscillates at $\Delta\omega$. The magnitude of the total field is given by,
\begin{equation}
|E_{atoms}|=\sqrt{E^2_{LO}+E^2_{Sig}+2{E_{LO}}{E_{Sig}}\cos\left(\Delta\omega t+\Delta\phi \right)}. \label{eq:mag}
\end{equation}
For weak fields where $E_{Sig}\ll E_{LO}$, (\ref{eq:mag}) becomes, 
\begin{equation}
\approx E_{LO}+E_{Sig}\cos\left(\Delta\omega t+\Delta\phi \right).
\label{eq:weakfield}
\end{equation}
The Rydberg atoms have a naturally different response to $E_{res}$ and $E_{mod}$. Since $E_{res}$ oscillates at $\omega_{LO}$ it is resonant with the Rydberg transition, where as $E_{mod}$ oscillates at a frequency that is well below the Rydberg transition frequency and results in a modulation of the EIT spectrum and thus the probe laser intensity on the photodiode (see Fig \ref{fig:setup}). The effect being the down conversion of the incident field $E_{2}$ from the base band RF frequency of  $\omega_{Sig}$ to an intermediate frequency (IF) of $f_{IF}$=$\Delta\omega/(2\pi)$ (see Fig.\ref{fig:flow}),
\begin{equation}
 E_{IF}=E_{LO}+E_{Sig}\cos\left(\Delta\omega t+\Delta\phi \right). \label{downconv}
\end{equation}
In this case the probe laser intensity on the photodiode varies at $f_{IF}$=90~kHz.  A detectable IF signal is produced even for $E_{sig}$ well below $E_{AT}$. Fig.~\ref{fig:IFsig} shows time domain plots of the IF signal out of the photodiode for various $E_{sig}$ levels. The 90~kHz modulation is easily seen as is the changing modulation amplitudes following the behavior of (\ref{downconv}). For the final stage of detection the output of the photodiode is passed to a lock-in amplifier with a reference set equal to the IF frequency, $f_{REF}$=$f_{IF}$. The lock-in output voltage ($V_{LI}$) is thus proportional to weak field, $V_{LI} \propto E_{Sig}$. 

Two identical source antennas (Narda 638 horns were used, however mentioning this product does not imply an endorsement by NIST, but only serves to clarify the equipment used) were used to produce $E_{LO}$ and $E_{Sig}$ fields. The antennas were placed 385 mm from the $^{133}$Cs vapor cell such that they were beyond the $2 \textbf{a}^2/\lambda_{RF}=305$~mm far field distance \cite{IEEEAntennasStandards}. Where $\textbf{a}=48.28$~mm is taken as the diagonal length of the antenna aperture and $\lambda_{RF}=15.286$~mm. Two separate RF signal generators synced via a 10~MHz reference were used to feed the two antennas at frequencies of $f_{LO}$=19.62600~GHz, and $f_{Sig}$=19.626090~GHz.  A calibrated power meter and vector network analyzer were used to account for cable loss from the RF signal generator and horn reflection coefficient and to determine the RF power at the horn antennas $P_{RF}$. For powers down to -70~dBm the power meter was used. For weak field generation $P_{RF}$ was <-70~dBm and thus well below the dynamic range of an RF power meter. To overcome this, the signal generator was operated within the range of the power meter from +10~to~-60~dBm and additional calibrated attenators were added providing up to $-111$~dB of additional loss. With this configuration accurate control of power levels could be achieved down to $\approx$~-180~dBm. 

\begin{figure}[htbp]
\includegraphics[width = 0.75\linewidth]{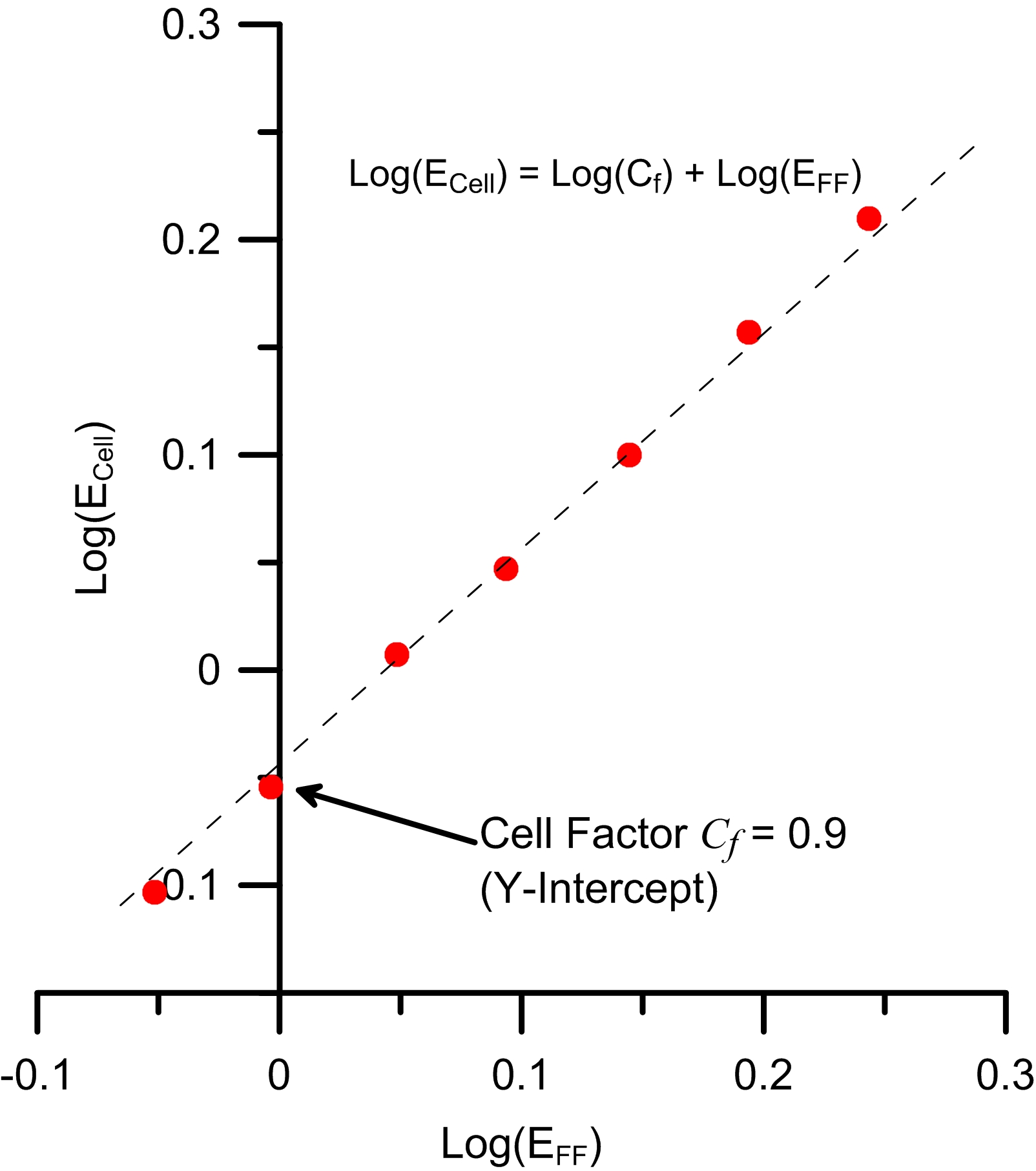}
\caption{\label{fig:cellfactor} Plot of $Log(E_{Cell})-v.s.-Log(E_{FF})$. $C_{f}$ is given by the Y-intercept.}
\end{figure}

\begin{figure}[htbp]
\includegraphics[width = 1.0\linewidth]{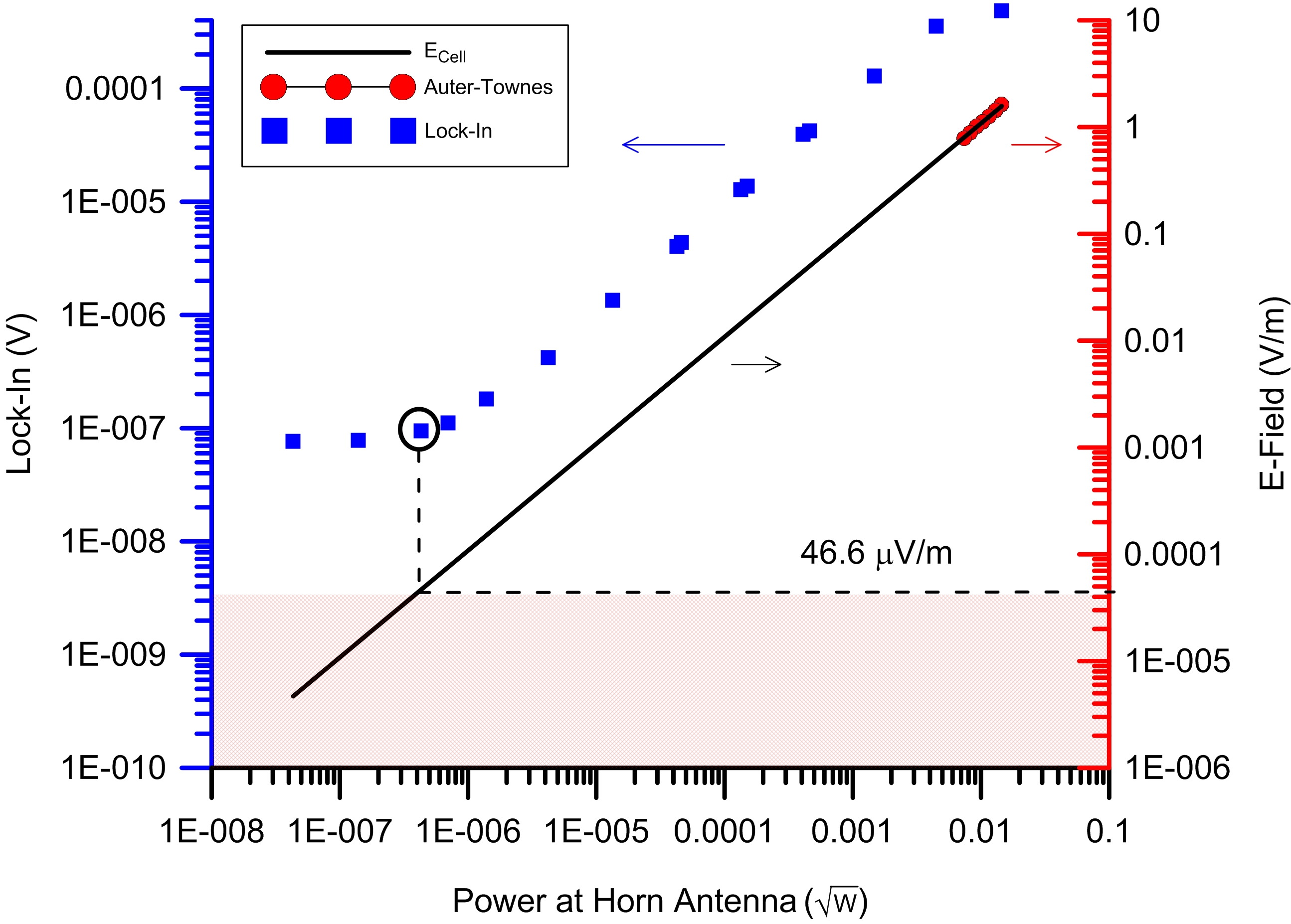}
\caption{\label{fig:weakfield} Detection plot for weak fields. (blue, left-axis) lock-in signal, (red, right-axis) AT splitting, and (line, right-axis) $E_{Cell}$ as a function of $\sqrt{P_{RF}}$. (dashed-line) Lock-in signal corresponding to lowest detectable E-field of 46~$\mu$V/m.}
\end{figure}

To accurately determine the E-field within the vapor cell for low RF powers into the horns, AT splitting was used to calibrate and correct errors imparted on the E-field due to the presence of the vapor cell. As has been shown in\cite{HollowaySubwave, Fan2015_vaporcellgeo, UncertaintiesSimons_2018} for an RF field incident on a vapor cell, scattering off of the glass walls can cause internal resonances and alter the E-field amplitude inside the vapor cell from that which would exist given the vapor cell were not there. The E-field at the horn-to-laser beam distance $R=$385~mm was calculated using \cite{IEEEAntennasStandards}$^,$\cite{balanis_AntennaC4} the far-field formula $E_{FF}=\sqrt{59.9585}\sqrt{ P_{RF} G}/{R}$ where the antenna gain is $G=$15.55~dB~$\pm$~0.4~dB. For a given distance $R$ and  RF frequency there is a fixed ratio of the E-field inside the vapor cell $E_{cell}$ to the E-field in the absence of vapor cell $E_{FF}$. This is given by the cell factor $C_f$=$E_{cell}/E_{FF}$. Calibration data for $E_{cell}$ was determined from the conventional AT splitting technique (\ref{eq:deltaf}) for a range of $P_{RF}$ strong enough to cause AT splitting. Cell factor calibration data comparing $E_{cell}$ and $E_{FF}$ is shown in Fig. \ref{fig:cellfactor}. Given the uncertainty in $G$,  power meter, and operating within the linear response\cite{Holloway2017systematicUnc} of the AT regime (\ref{eq:deltaf}), weak E-fields detected by the Rydberg mixer could be known for a given $P_{RF}$ to within an estimated uncertainty of $\pm$~$\%$~5.  For the configuration used here $C_f$=0.90 and thus for a given $P_{RF}$, 
\begin{equation}
E_{Cell}=\frac{0.90 \sqrt{59.9585}\sqrt{ P_{RF} G}}{R}
\label{eq:farfield}
\end{equation}
Weak E-field data (blue squares) are plotted in Fig.~\ref{fig:weakfield} for lock-in amplifier output voltage-vs-$\sqrt {P_{RF}}$ along with the corresponding E-field strength. For these data a 3~s time constant and 24~dB/octave low pass filter slope was used. Each data point is comprised of 3 data averages with standard deviation error of $\%$~5. As $P_{RF}$ approaches powers $<$~-100~dBm the lock-in signal approaches the noise floor which shows up by the flattening out of the data curve. Also shown in Fig. \ref{fig:weakfield} are the higher E-fields that were used for cell factor calibration and acquired from AT splitting . These data (red circles) follow the linear behavior predicted by equations (\ref{eq:deltaf}) \& (\ref{eq:farfield}). The weak E-field data remains linear up until $E_{AT}$ is reached. The cross over between the weak field regime and AT regime shows up as a roll off of the weak field data near $E_{AT}$.  This roll off is due to the EIT peak center frequency shifting away from the probe laser frequency as AT splitting begins to take place. The weakest detectable E-field is taken as the value at where the lock-in voltage curves to the noise floor.  This corresponds to $\approx 46$~$\mu$V/m.

\begin{figure}[htbp]
\includegraphics[width = 1.0\linewidth]{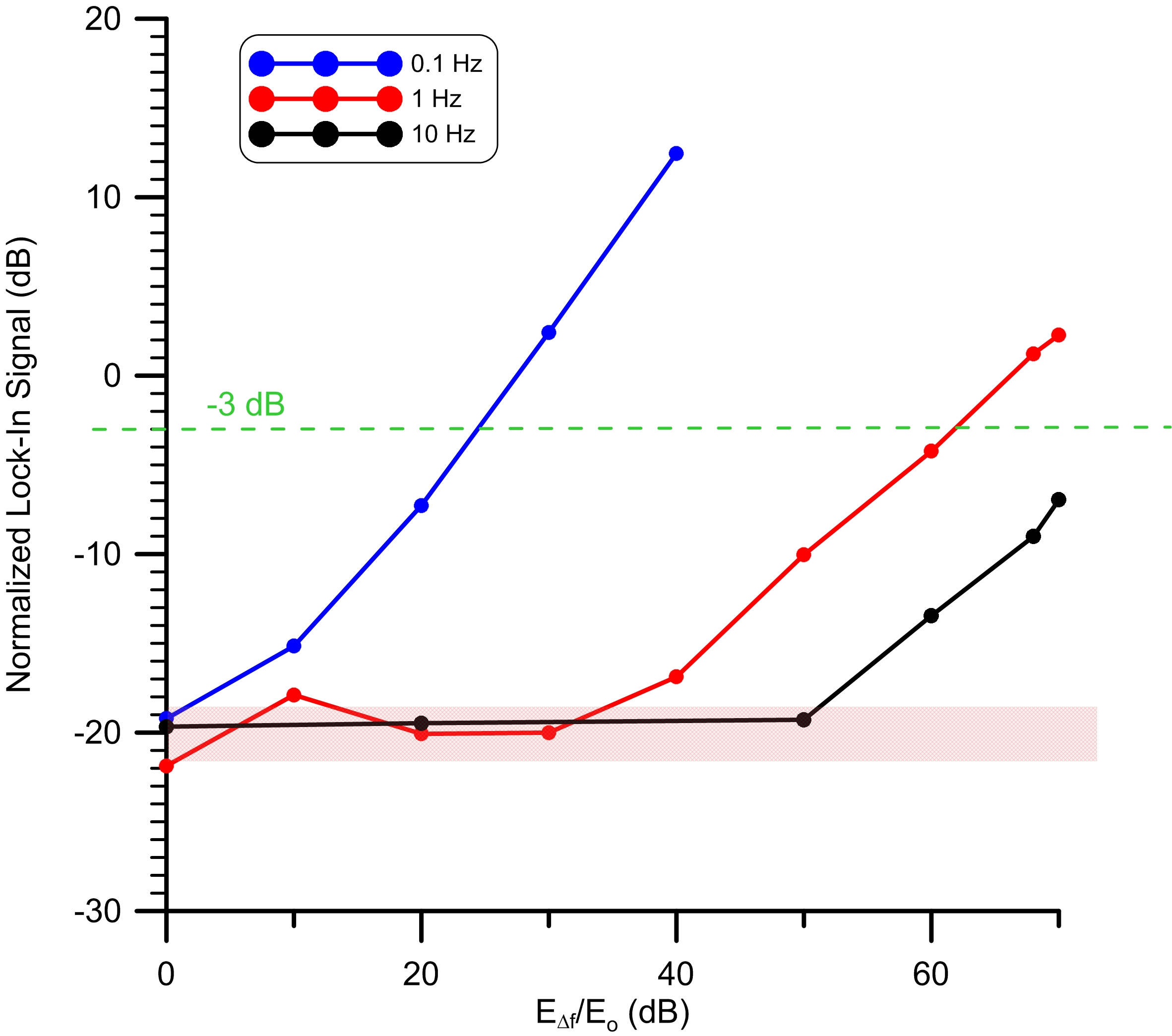}
\caption{\label{fig:outof}Isolation of neighboring signals for various E-field strengths relative to $E_{o}$ and for $\Delta$f= 0.1~Hz, 1~Hz, and 10~Hz. Lock-in signal is normalized to that produced by $E_{o}$=181~$\mu$V/m. Signals below $-3$~dB level are considered to be isolated. Noise floor around $-20$~dB is shown by red region. }
\end{figure}

Another aspect of the Rydberg mixer is its ability to isolate and discriminate between signals of differing RF frequencies with a frequency resolution orders of magnitude finer than the response bandwidth of the Rydberg transition. As was shown in \cite{Simons_2016_Freq_Detuning}, through the generalized Rabi frequency, RF E-fields that are off-resonance with the Rydberg transition will still affect the EIT spectrum over a large continuum of frequencies of hundreds of MHz. For an RF frequency detuning of $\delta_{RF}$, and on-resonance Rabi frequency of $\Omega_{o}$, the generalized Rabi frequency becomes, $\Omega'=\sqrt{\Omega_{o}^2+\delta_{RF}^2}$. For example in the AT regime, splitting will still occur for off-resonance E-fields for a large range of $\delta_{RF}$, where now the splitting $\Delta f_{AT}\rightarrow \Omega'/(2\pi)$.  As such, discriminating between E-fields of different RF frequencies through purely observing the EIT spectrum becomes difficult and ambiguous. The Rydberg atom mixer provides a means to overcome this so that E-fields differing in frequency by as little as 1~Hz can be discriminated. For this, the lock-in amplifier is tuned to the desired IF frequency corresponding to the desired down converted RF frequency. Simply tuning $f_{REF}$ allows for signals at different RF frequencies to be discriminated and isolated.\\
\indent We demonstrate this and examine the leakage in the lock-in signal for E-fields at neighboring frequencies and various strengths relative to the "in-tune" E-field. First, an in-tune IF signal was produced where the RF signal generator power was set to roughly middle of range at $P_{RF}$=-40~dBm and $f_{IF}$=90~kHz. This signal we denote as $E_{o}$=181~$\mu$V/m. The lock-in reference was also tuned to $f_{REF}$=90~kHz, and a time constant of 3~s, giving a cut off frequency of $f_{c}$=0.33~Hz. Three other signals denoted as $E_{\Delta f}$ that were out of tune by $\Delta f$=0.1~Hz, $\Delta f$=1~Hz, $\Delta f$=10~Hz were also produced. For these three signals $P_{RF}$ was then varied such that $E_{\Delta f}/E_{o}$ ranged from 0~dB to greater than 60~dB. Fig. \ref{fig:outof} shows a plot of the lock-in output for the three detuned signals normalized to the level produced by $E_{o}$. The lock-in noise floor is depicted as well. As can be seen there is a range of relative strengths for each detuned signal where the lock-in signal is at the noise floor and then rises up to equal the level of $E_{o}$. All three detunings show maximum isolation when equal to $E_{\Delta f}/E_{o}=0$~dB.  Where even for sub-Hz detuning of $\Delta f$=0.1~Hz, $E_{\Delta f}$ does not rise above the noise floor. The isolation threshold in dB for each detuning is taken for the value of  $E_{\Delta f}/E_{o}$ that crosses -3~dB level of the lock-in signal. Isolation degrades more quickly for smaller detunings for $E_{\Delta f}/E_{o}>$1. For a detuning of $\Delta f$=1~Hz the -3~dB crossing happens for $E_{\Delta f}/E_{o}\approx 60$~dB.  

This work shows E-field strengths -84~dB below the AT limit $E_{AT}$ can be detected using the Rydberg atom mixer\cite{SimonsAtomMixer2019}. This is better than an order of magnitude improvement in the minimum detectable E-field compared to previously reported techniques ($\approx$~46~$\mu$V/m~$\pm$~2~$\mu$V/m as opposed to 800~$\mu$V/m \cite{Sedlacek2012Nature}).  Furthermore, the Rydberg atom mixer allows specific RF frequencies to be selected, isolated and rejected with resolution better than 1~Hz. This is a $\backsim 10^8$ improvement in RF frequency resolution over that provided by the frequency bandwidth\cite{Simons_2016_Freq_Detuning} of the Rydberg transition alone. These attributes along with the ability to measure phase\cite{SimonsAtomMixer2019}, and polarization\cite{Sedlacek2013AtomBasedVector} allow for the development of a quantum-based sensor to fully characterize the RF E-field in one compact vapor cell. 


%

\end{document}